\documentclass[10pt,twocolumn,letterpaper]{article}

\usepackage{cvpr}
\usepackage{times}
\usepackage{epsfig}
\usepackage{graphicx}
\usepackage{amsmath}
\usepackage{amssymb}

\usepackage{bm}
\usepackage{subcaption}

\usepackage{algorithm,algpseudocode}
\usepackage{multirow}
\usepackage{booktabs}
\usepackage{xcolor}

\usepackage[pagebackref=true,breaklinks=true,letterpaper=true,colorlinks,bookmarks=false]{hyperref}

\cvprfinalcopy 

\def\cvprPaperID{****} 

\begin{document}
	
	\title{NTIRE 2020 Challenge on Video Quality Mapping: Methods and Results}
	
	\author{Dario Fuoli \and
 Zhiwu Huang \and Martin Danelljan \and Radu Timofte \and Hua Wang \and Longcun Jin \and Dewei Su \and Jing Liu \and Jaehoon Lee \and Michal Kudelski \and Lukasz Bala \and Dmitry Hrybov \and Marcin Mozejko \and Muchen Li \and Siyao Li \and Bo Pang \and Cewu Lu \and Chao Li \and Dongliang He \and Fu Li \and Shilei Wen}
	
	\maketitle
\thispagestyle{empty}

	\begin{abstract}
		This paper reviews the NTIRE 2020 challenge on video quality mapping (VQM), which addresses the issues of quality mapping from source video domain to target video domain. The challenge includes both a supervised track (track 1) and a weakly-supervised track (track 2) for two benchmark datasets. In particular, track 1 offers a new Internet video benchmark, requiring algorithms to learn the map from more compressed videos to less compressed videos in a supervised training manner. In track 2, algorithms are required to learn the quality mapping from one device to another when their quality varies substantially and weakly-aligned video pairs are available. For track 1, in total 7 teams competed in the final test phase, demonstrating novel and effective solutions to the problem. For track 2, some existing methods are evaluated, showing promising solutions to the weakly-supervised video quality mapping problem. 
	\end{abstract}
{\let\thefootnote\relax\footnotetext{%
\hspace{-5mm}Dario Fuoli (\texttt{dario.fuoli@vision.ee.ethz.ch}), Zhiwu Huang, Martin Danelljan, and Radu Timofte at ETH Z\"urich are the NTIRE 2020 challenge organizers. The other authors participated in the challenge.
Appendix A contains the authors' teams and affiliations.\\
\url{https://data.vision.ee.ethz.ch/cvl/ntire20/}
}}
	
\section{Introduction}
Human captured and transmitted videos often suffer from various quality issues. For instance, despite the incredible development of current smartphone or depth cameras, compact sensors and lenses still make DSLR-quality unattainable for them. Due to bandwidth limit over internet, videos have to be compressed for easier transmission. The compressed videos inevitably suffer from compression artifacts. Therefore, quality enhancement over such videos are highly in demand. 

The challenge aims at pushing competing methods into effective and efficient solutions to the newly emerging video quality mapping (VQM) tasks. Following \cite{kim2019vid3oc}, two tracks are studied in this challenge. Track 1 is configured to the task of fully-supervised video quality mapping between more compressed videos to less compressed videos collected from the Internet, while track 2 are designed for the weakly-supervised video quality mapping from a ZED camera to a Canon 5D Mark IV camera. Competing methods are evaluated with the most prominent metrics in the field, \ie, Peak Signal-to-Noise Ratio (PSNR) and structural similarity index (SSIM). 

Since PSNR and SSIM are not always well correlated with human perception of quality, we also consider to leverage perceptual measures, such as the Learned Perceptual Image Patch Similarity (LPIPS)~\cite{zhang2018perceptual} metric as well as mean opinion scores (MOS), which aim to evaluate the quality of the outputs according to human visual perception.


This challenge is one of the NTIRE 2020 associated challenges on: deblurring~\cite{nah2020ntire}, nonhomogeneous dehazing~\cite{ancuti2020ntire}, perceptual extreme super-resolution~\cite{zhang2020ntire}, video quality mapping (this paper), real image denoising~\cite{abdelhamed2020ntire}, real-world super-resolution~\cite{lugmayr2020ntire}, spectral reconstruction from RGB image~\cite{arad2020ntire} and demoireing~\cite{yuan2020demoireing}.
	\section{Related Work}

\noindent\textbf{Quality Enhancement on Compressed Videos} aims to eliminate visual artifacts of compressed videos, which are transmitted over the bandwidth-limited Internet and often suffers from compression artifacts. There are emerging several algorithms like \cite{dai2017convolutional,wang2017novel,yang2018multi}, which generally employ the original (uncompressed or less compressed) videos for full supervision on video quality map learning. For instance,
\cite{wang2017novel} proposes an Auto-Decoder to learn the non-linear mapping from the decoded video to the original one, such that the artifacts can be removed and  details can be enhanced on compressed videos. \cite{dai2017convolutional} suggests a post-processing algorithm for artifact reduction on compressed videos. Based on the observation that High in Efficiency Video Coding (HEVC) adopts variable block size transform, the suggested algorithm integrates variable filter size into convolutional networks for better reduction of the quantization error. To take advantage of the information available in the neighboring frames, \cite{yang2018multi} proposes a deep network to take both current frame and its adjacent high-quality frames into account for better enhancement on compressed videos.

\noindent\textbf{Video Super-Resolution (VSR)} methods are used as well to enhance the texture quality of videos. The requirements for VSR and VQM are similar. It is important to enforce temporally consistent transitions between enhanced frames and to accumulate information over time, which is a fundamental difference to single image enhancement methods.
Most deep learning based methods adopt the idea of concatenating adjacent frames with explicit motion compensation in order to leverage temporal information \cite{kappeler, tao, caballero}. A more recent method \cite{duf} successfully explores the application of 3D convolutions as a natural extension for video data, without explicit motion compensation.
In contrast to single image enhancers, many applications for video require real-time performance. Therefore, efficient algorithms for video processing are in high demand. Temporal information can be very efficiently aggregated with recurrent neural networks (RNN) which are developed in \cite{frvsr, rlsp}. For instance, \cite{frvsr} efficiently warps the previous high-resolution output towards the current frame according to optical flow. In \cite{rlsp}, runtimes are further improved by propagating an additional hidden state, which handles implicit processing of temporal information without explicit motion compensation.
Perceptual improvements over fully-supervised VSR methods are realized with generative adversarial networks (GAN) by \cite{tecogan} and \cite{photorealvsr}.

\noindent\textbf{Quality Enhancement on Device Captured Videos} aims at enhancing the perceived quality of videos taken by devices, which includes enhancements like increasing color vividness, boosting contrast, sharpening up textures, etc.
However, the major issue of enhancing such videos is the extreme challenge of collecting well-aligned training data, i.e., input and target videos that are aligned in both the spatial and the temporal domain. 
A few approaches address this problem using  reinforcement learning based techniques like \cite{hu2018exposure,park2018distort,kosugi2019unpaired}, which aims at creating pseudo input-retouched pairs by applying retouching operations sequentially.

Another direction is to develop Generative Adversarial Network (GAN) based methods for this task. For example, \cite{chen2018deep} proposes a method for image enhancement by learning from unpaired photographs. The method learns an enhancing map from a set of low-quality photos to a set of high-quality photographs using the GAN technique  \cite{goodfellow2014generative}, which has proven to be good at learning real data distributions. Similarly, \cite{ignatov2018wespe} leverages the GAN technique to learn the distribution of separate visual elements (i.e., color and texture) of images, such that the low-quality images can be mapped easier to the high-quality image domain which is encoded with more vivid colors and more sharpened textures. More recently, \cite{huang2019divide} suggests a divide-and-conquer adversarial learning method to further decompose the photo enhancement problem into multiple sub-problems. Such sub-problems are divided hierarchically: 1) a perception-based division for learning on additive and multiplicative components, 2) a frequency-based division in the GAN context for learning on the low- and high-frequency based distributions, and 3) a dimension-based division for factorization of high-dimensional distributions. To further smooth the temporal semantics during the enhancement,  an efficient recurrent design of the GAN model is introduced. To the best of our knowledge, except for \cite{huang2019divide}, there are very few works specially for weakly-supervised video enhancement.
    \begin{figure*}[t]
    \centering
	\includegraphics[width=0.85\textwidth]{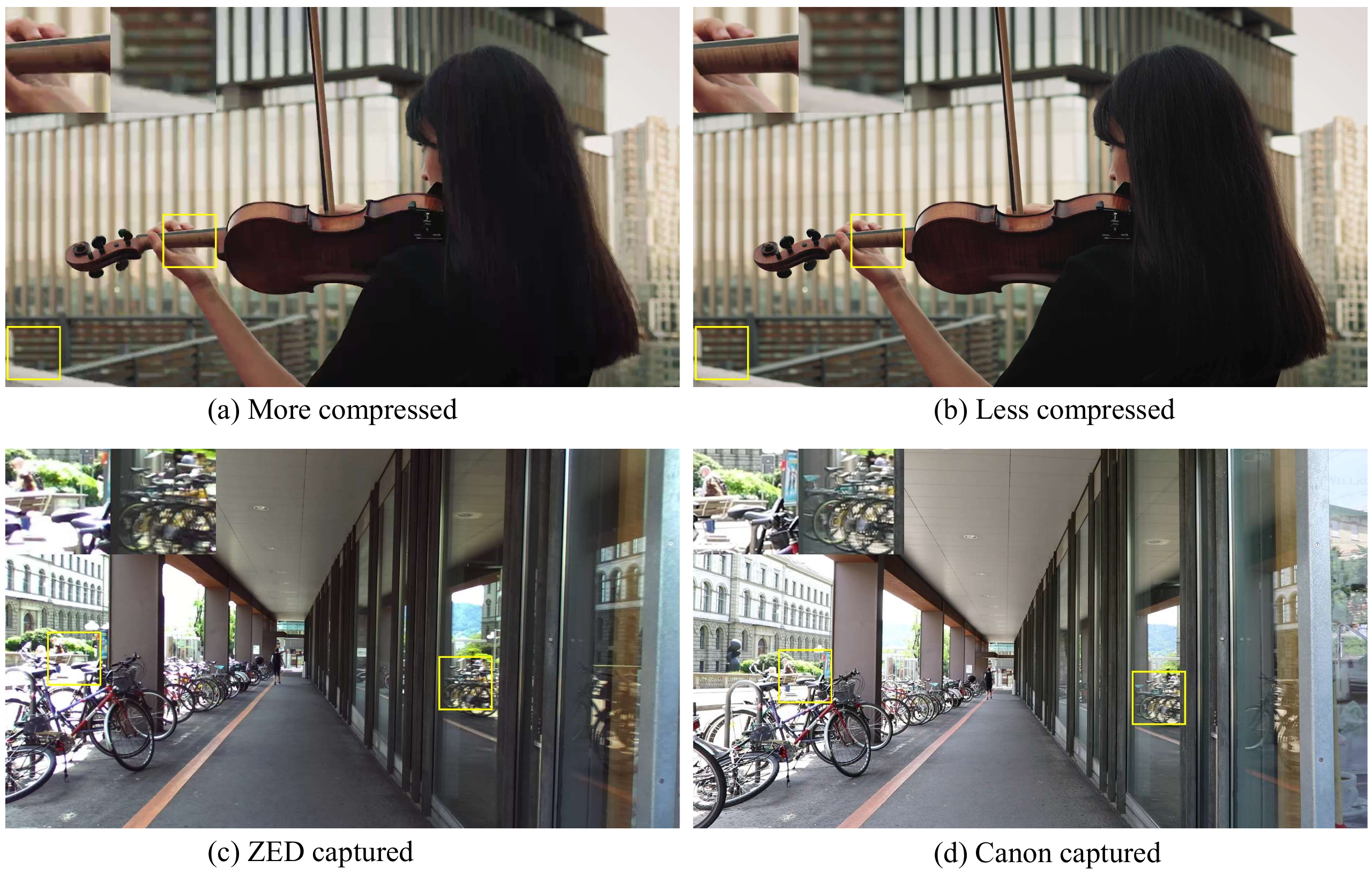}%
	\caption{Track 1 (a)-(b): quality mapping from more compressed (a) to less compressed (b) videos which are well aligned. Track 2 (c)-(d): quality mapping from low-quality videos captured by a ZED camera (c) to high-quality Canon DSLR videos (d), which are roughly aligned.}%
	\label{fig:track1_track2_example}\vspace{1mm}
\end{figure*}

\section{Challenge Setup}

\subsection{Track 1: Supervised VQM}

For this track, we introduce the IntVid dataset \cite{kim2019vid3oc}. It consists of videos downloaded from Internet websites. The  collected videos cover 12 diverse scenarios: city, coffee, fashion, food, lifestyle, music/dance, narrative, nature, sports, talk, technique and transport. The resolution of the crawled videos is mostly 1920$\times$1080. Their duration varies from 8 seconds to 360 seconds with frame rates in the range of 23.98-25.00 FPS. 

As most of the collected videos consist of changing scenes, a popular scene detection tool named PySceneDetect\footnote{\url{https://pyscenedetect.readthedocs.io}} is used to split the videos into three separate sets of clips for training, validation and test respectively. 
In particular, most of the resulting video clips are selected such that the majority of the original video content is employed for training. For the validation and test video clips, the video length is fixed to 4 seconds containing 120 frames, which are saved as PNG image files. 

Due to the bandwidth limit of Internet, video compression techniques are often applied to reduce the coding bit-rate. Inspired by this, \cite{kim2019vid3oc} applied the standard video coding system H.264 to compress the collected videos. As a result, a total of 60 paired compressed and uncompressed videos are generated for training, and 32 paired compressed/uncompressed clips are produced for validation and testing. One example for track 1 is shown in Fig.\ref{fig:track1_track2_example} (a)-(b).

\subsection{Track 2: Weakly-Supervised VQM}

For this track, we employ the Vid3oC dataset \cite{kim2019vid3oc}, which records videos with a rig containing three cameras. 
In particular, we use the Canon 5D Mark IV DSLR camera to serve as a high-quality reference, while utilizing the ZED camera, which additionally records depth information, to provide sequences of the same scene with a significantly lower video quality level. 
As the track focuses on the RGB-based visual quality mapping, we remove the depth information from the ZED camera. Using the two cameras, videos are recorded in the area in and around Zurich, Switzerland during the summer months. The locations and scenes are carefully chosen to ensure variety in content, appearance, and the dynamic nature. The length of each recording is between 30 and 60 seconds. Videos are captured in 30 FPS, using the highest resolution (i.e., 1920$\times$1080) available at that frame rate.

In \cite{kim2019vid3oc}, the recorded videos are split into a training set of 50 weakly-paired videos, together with a validation and test set of 16 videos each. 
For all sets, a rough temporal alignment is performed based on the visual recording of a digital clock, which is captured by both cameras in the beginning of each video. 
The training videos are then trimmed down to 25-50 seconds by removing the first few seconds (which include the timer) and encoded with H.264. For each video in the validation and test set, a 4-second interval is selected. Each of such small video clips contains 120 frames, which are stored as individual PNG image files. Fig.\ref{fig:track1_track2_example} (c)-(d) illustrates one example for track 2.

\subsection{Evaluation Protocol}

\noindent\textbf{Validation phase}: During the validation phase, the source domain videos for the validation set were provided on CodaLab. While the participants had no direct access to the validation ground truth, they could get feedback through the online server on CodaLab. Due to the storage limits on the servers, the participants could only submit a subset of frames for the online evaluation. PSNR and SSIM were reported for both tracks, even though track 2 only has weakly-aligned targets. The participants were allowed to make 10 submissions per day, and 20 submissions in total for the whole validation phase.

\noindent\textbf{Test phase}: In the test phase, participants are expected to submit their final results to the CodaLab test server. Compared to the validation phase, no feedback was given in terms of PSNR/SSIM to prevent comparisons with other teams and overfitting to the test data. By the deadline, the participants were required to provide the full set of frames, from which the final results were obtained.

\section{Challenge Teams and Methods}

In total 7 teams submitted their solutions to track 1. One team asked to anonymize their team name and references, since they found out to be using inappropriate extra-data for training after the test phase submission deadline. No submissions were made for track 2.   

\begin{figure}[t]
    \centering
	\includegraphics[width=0.45\textwidth]{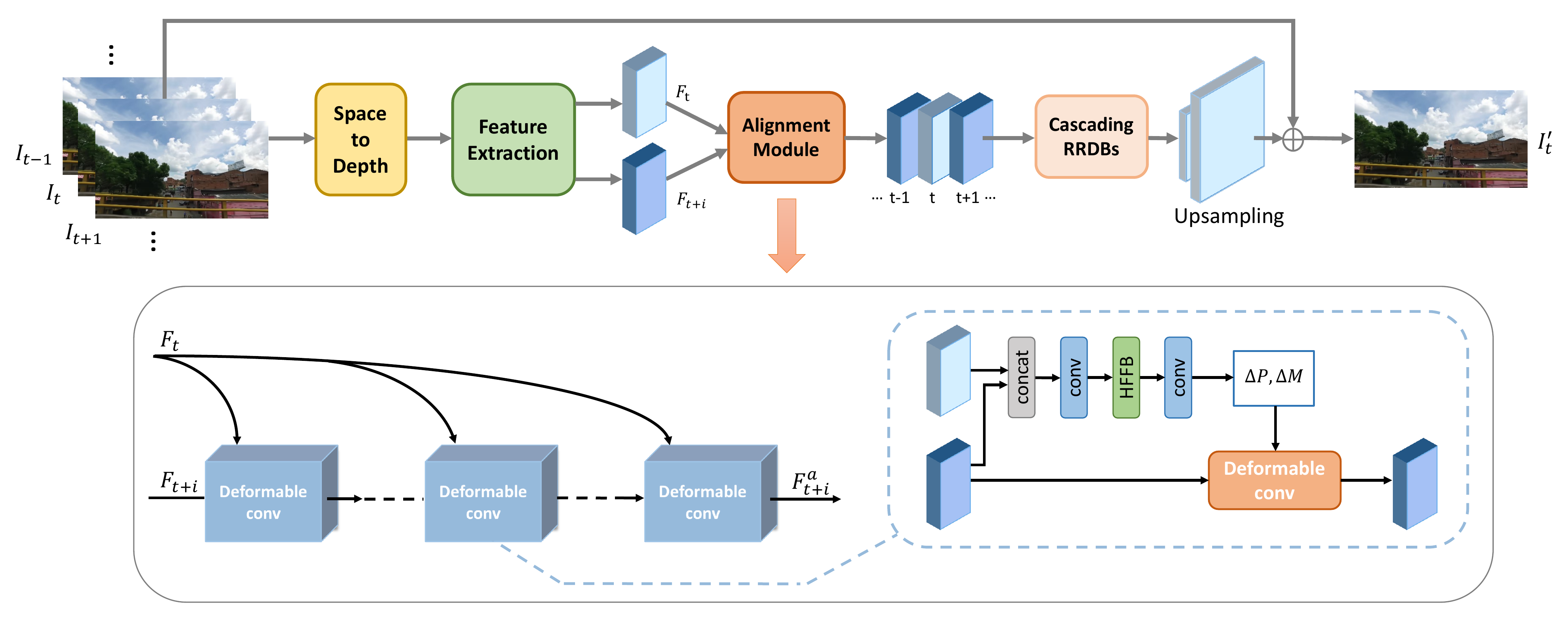}%
	\caption{Illustration of the network design suggested by team GTQ.}%
	\label{fig:GTQ_network}
	\vspace{-5mm}
\end{figure}
\begin{figure}[t]
    \centering
	\includegraphics[width=0.45\textwidth]{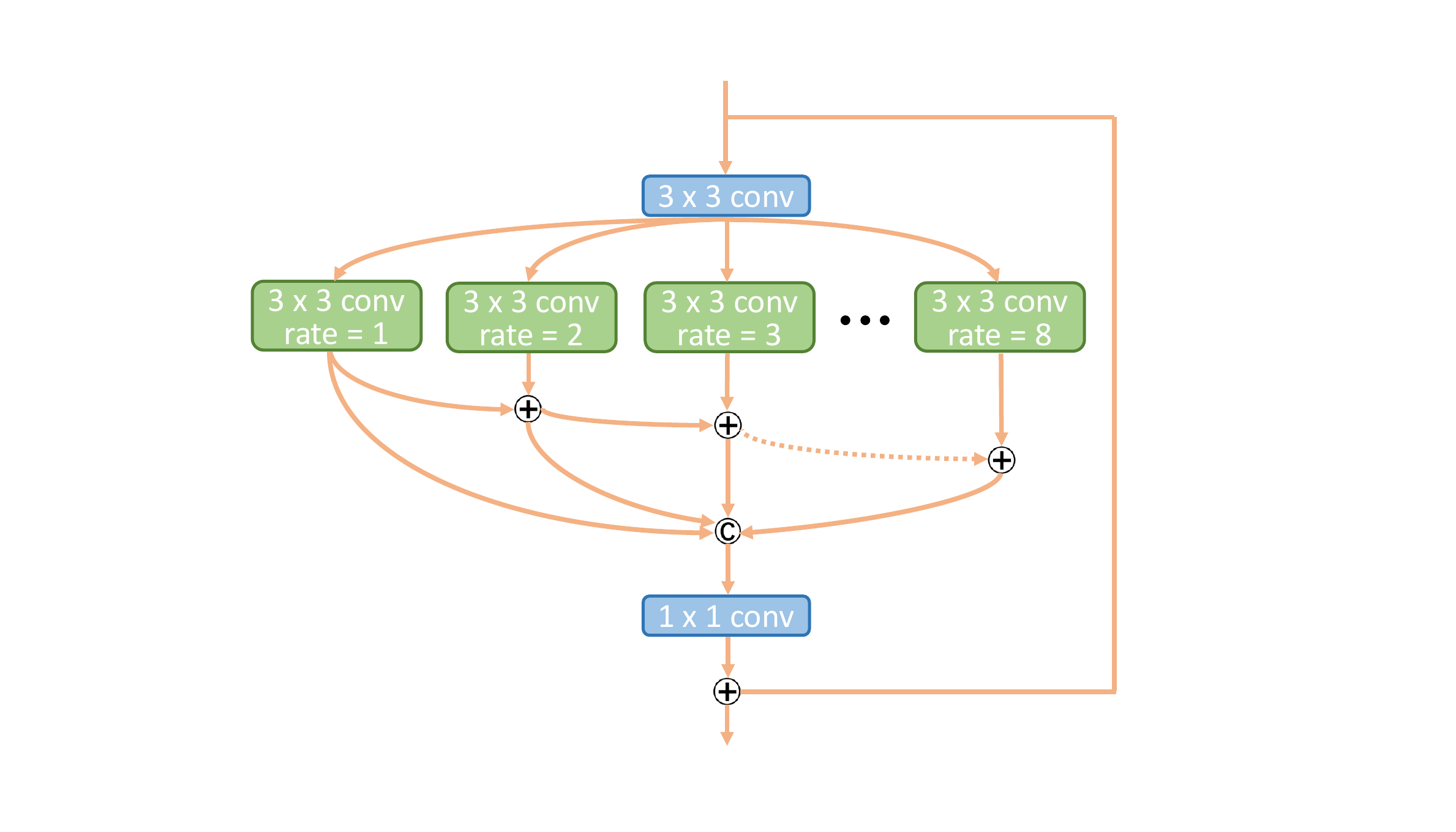}
	\vspace{-5mm}
	\caption{Illustration of the hierarchical feature fusion block (HFFB) suggested by team GTQ.}%
	\label{fig:GTQ_hffb}
	\vspace{1mm}
\end{figure}


\subsection{GTQ team}

The team proposes a modified deformable convolution network to achieve high quality video mapping as shown in Fig.~\ref{fig:GTQ_network}. The framework first down-samples the input frames with scale factor 4 through a space to depth shuffling operation. Then, the extracted features pass through an alignment module which applies a cascade of deformable convolutions \cite{zhu2019deformable} to perform implicit motion compensation. In the alignment module, the team takes advantage of hierarchical feature fusion blocks (HFFB) \cite{hui2019progressive} to predict more precise offset and modulation scalars used in deformable convolutions. As shown in Fig.~\ref{fig:GTQ_hffb}, HFFB introduces a spatial pyramid of dilated convolutions to effectively enlarge the receptive field with relatively low computational cost, which contributes to dealing with complicated and large motions between frames. After the alignment operation, the features are concatenated and fed into stacked residual in residual dense blocks (RRDB) \cite{wang2018esrgan} to reconstruct high quality frames.

\begin{figure}[t]
		\centering
		\includegraphics[scale = 0.3]{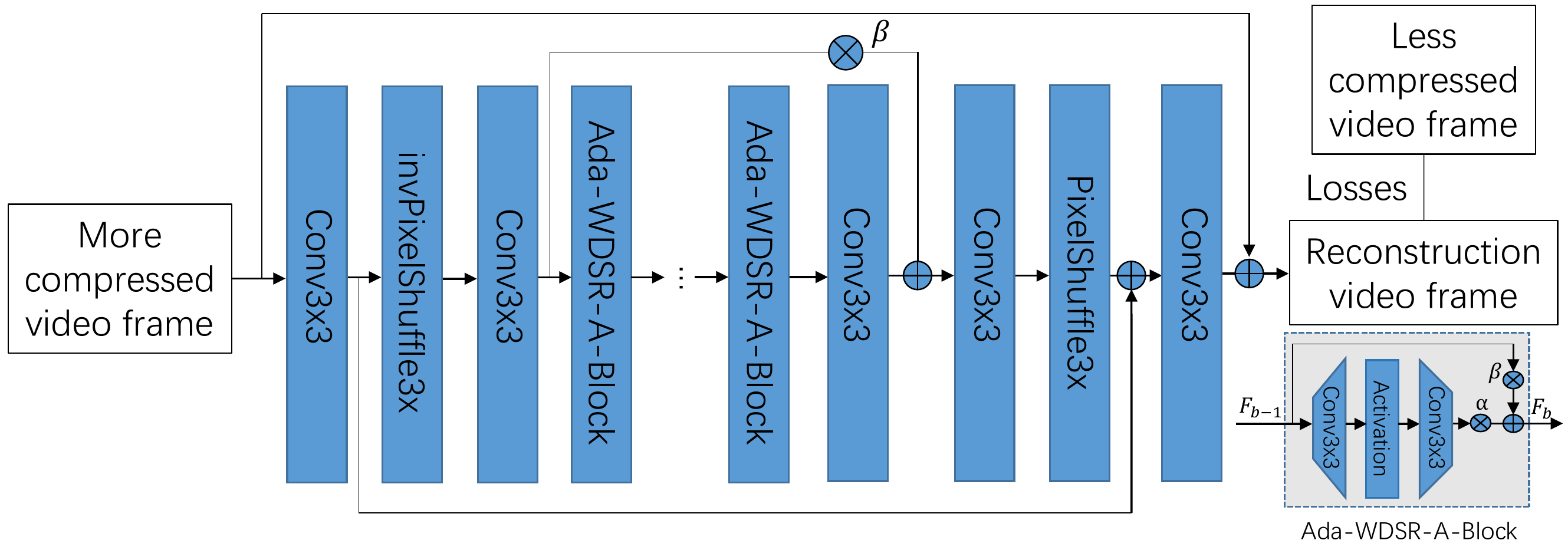}
		\caption{The network architecture of the proposed C2CNet by team ECNU. }
		\label{fig1}
\end{figure}

\subsection{ECNU team} 

Team ECNU proposes a Compression to Compression Network (C2CNet). The input to C2CNet is a more compressed video frame and the ground truth is a less compressed video frame. As shown in Fig. \ref{fig1}, C2CNet is composed of a head $3\times3$ convolutional layer, a de-sub-pixel convolutional layer composed of an inverse pixel-shuffle layer and a $3\times3$ convolution, a non-linear feature mapping module composed of 64 Adaptive WDSR-A-Blocks, a $3\times3$ convolution and a short skip connection with residual scaling $\beta$=0.2, an upsampling skip connection, a sub-pixel convolutional layer composed of a $3\times3$ convolution and a pixel-shuffle layer, a global skip connection and a tail $3\times3$ convolution. The number of channels for C2CNet is 128. The Adaptive WDSR-A-Block is composed of 64, 256 and 64 channels. The Adaptive WDSR-A-Block is modified from a WDSR-A-Block \cite{yu2018wide}, by adding learnable weight $\alpha$ (initialized with 1) for body scaling and learnable weight $\beta$ (initialized with $0.2$) for residual scaling. Each $3\times3$ convolution is followed by a weight normalization layer (omitted in Fig.~\ref{fig1}).

 \begin{figure}[t]
        \centering
    \begin{subfigure}{0.5\textwidth}
        \centering
        \includegraphics[width=0.5\linewidth]{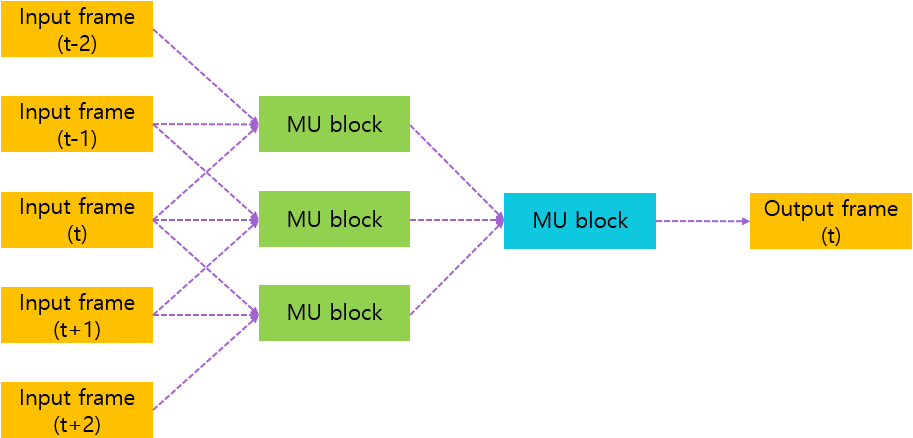}
        \caption{}
        \label{fig:GIL_1}
    \end{subfigure}
    \begin{subfigure}{0.5\textwidth}
        \centering
        \includegraphics[width=0.5\linewidth]{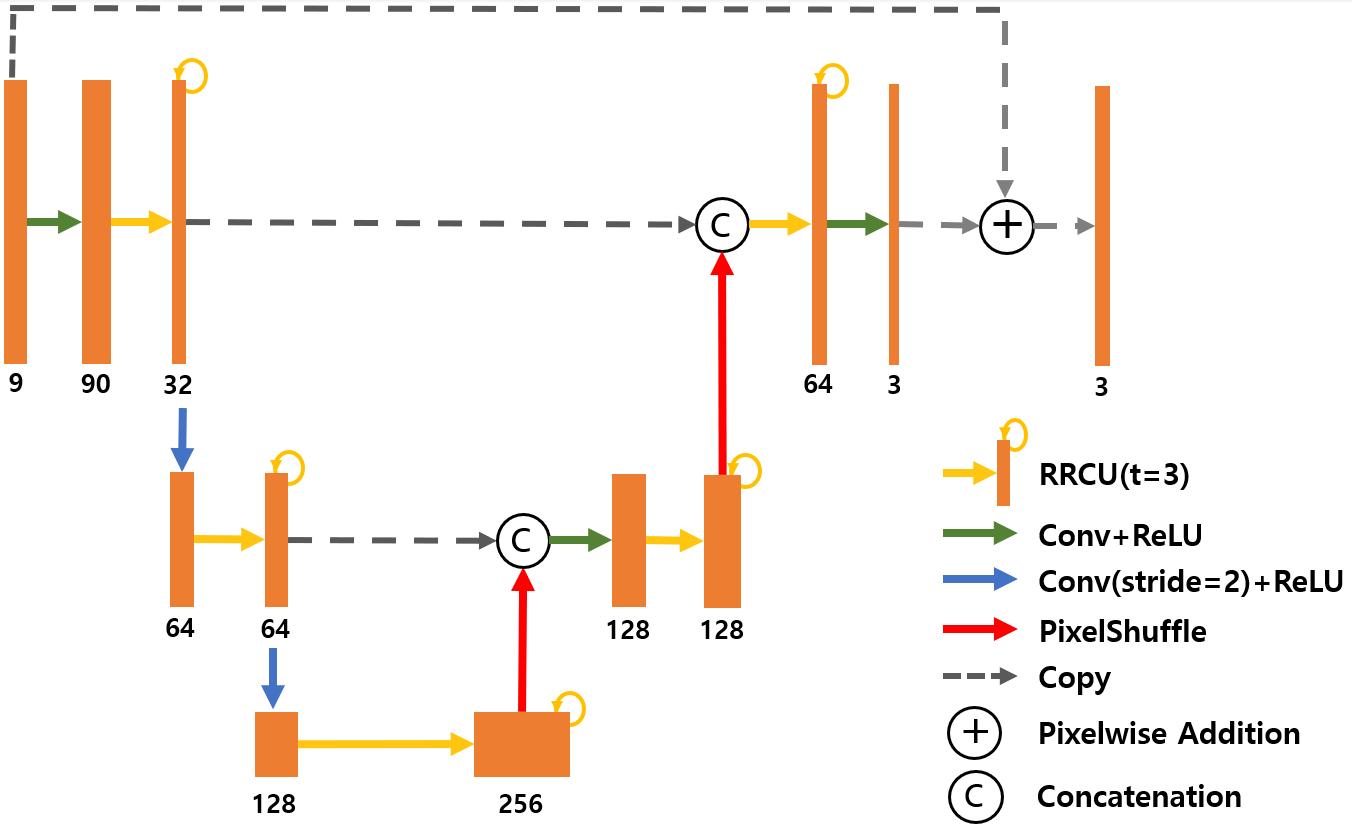}
        \caption{}
        \label{fig:GIL_2}
    \end{subfigure}
    \begin{subfigure}{0.5\textwidth}
        \centering
        \includegraphics[width=0.5\linewidth]{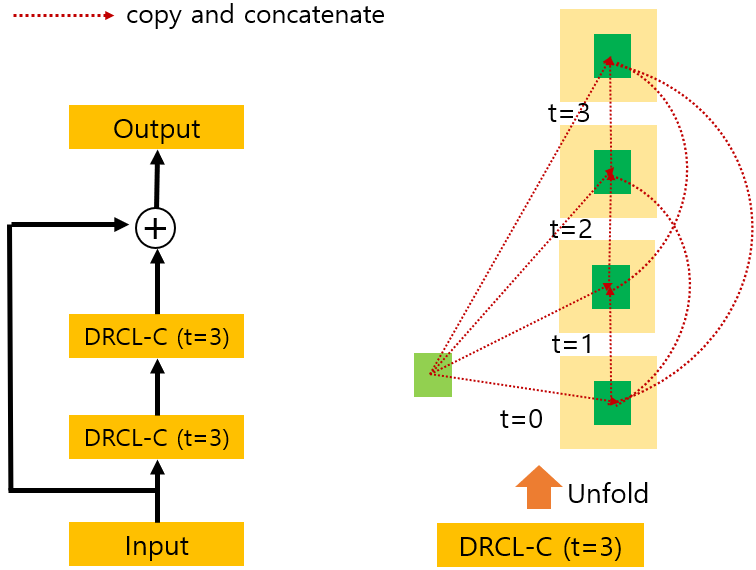}
        \caption{}
        \label{fig:GIL_3}
    \end{subfigure}
    
    \caption{Architecture of team GIL's model. (a) Overall Network architecture. (b) MU block. (c) RRCU (t=3) at the left and Unfolded DRCL-C (t=3) at the right.}
    \label{}
    \end{figure}

\subsection{GIL team} 

The team employs a network with two-stage architecture proposed in FastDVDnet~\cite{tassano2019fastdvdnet} and is shown in Fig.~\ref{fig:GIL_1}. It takes five consecutive frames as an input and generates a restored central frame. Three MU blocks in the first stage (shown in green) share parameters. Each MU block is a modified U-Net~\cite{ronneberger2015u} shown in Fig.~\ref{fig:GIL_2}. It uses a convolutional layer with stride=2 for down-sampling and a pixel-shuffle \cite{shi2016real} layer for up-sampling. It features a skip connection for global residual learning and contains several RRCUs (recurrent residual convolutional unit) inspired from R2U-Net \cite{alom2018nuclei}. Each RRCU consists of two DRCL-C (dense recurrent convolutional layer-concatenate) and a skip connection for residual learning. Figures of RRCU and DRCL-C are shown in Fig.~\ref{fig:GIL_3}. States of the DRCL-C change over discrete time steps and the maximum time step is limited to 3. The DRCL-C is different from a standard RCL (recurrent convolutional layer) \cite{liang2015recurrent}. It reuses previous features by concatenating them \cite{huang2017densely}.  A convolutional layer with 1x1 filters is used after every concatenation in DRCL-C to make the number of channels constant. The network has approximately 3.6 million parameters.

\begin{figure}[t]
    \centering
    \includegraphics[scale=0.25]{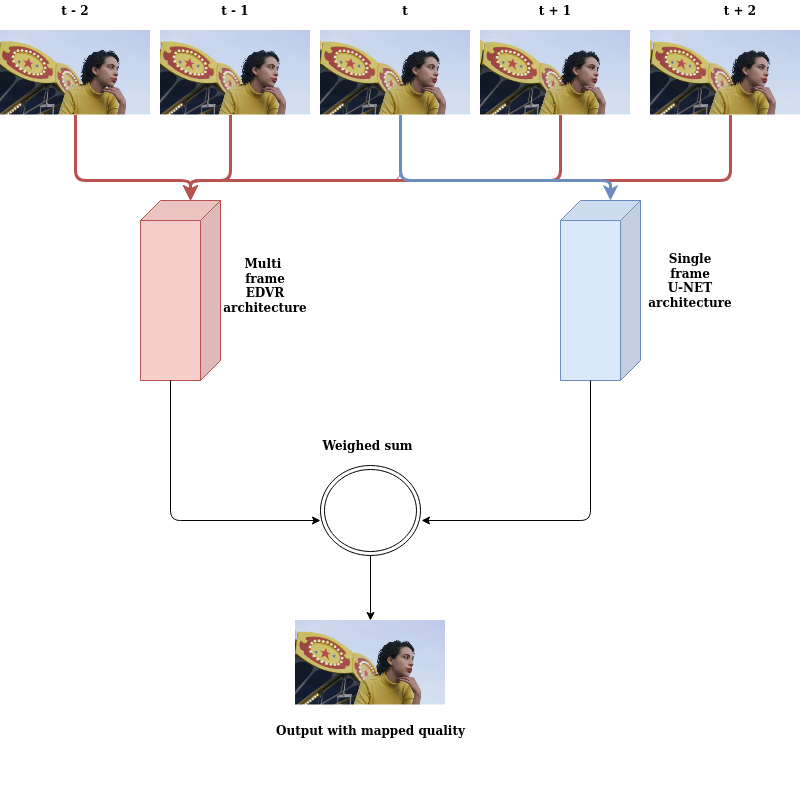}
    \vspace{-5mm}
    \caption{Schematic representation of team TCL's approach.}
    \label{fig:tcl_network}
    
\end{figure}

\subsection{TCL team} 

The team uses a pyramidal architecture with deformable convolutions and spatio-temporal attention based on the work of \cite{wang2019edvr} along with a single-frame U-Net \cite{ronneberger2015u}. The overview of the method is illustrated in Fig.~\ref{fig:tcl_network}. By combining these two methods, the local frame structure is preserved with the usage of U-Net and additional information from neighboring frames along with motion compensation, mostly by exploiting the PCD module from \cite{wang2019edvr}, is used to enhance output quality. Both networks are trained separately and the final result is obtained by a weighted sum with weight parameter $\beta$ found by grid search, which is validated on a hold-out set from the training frames.

\begin{figure}[t]
\centering
		\includegraphics[width = 0.95\linewidth]{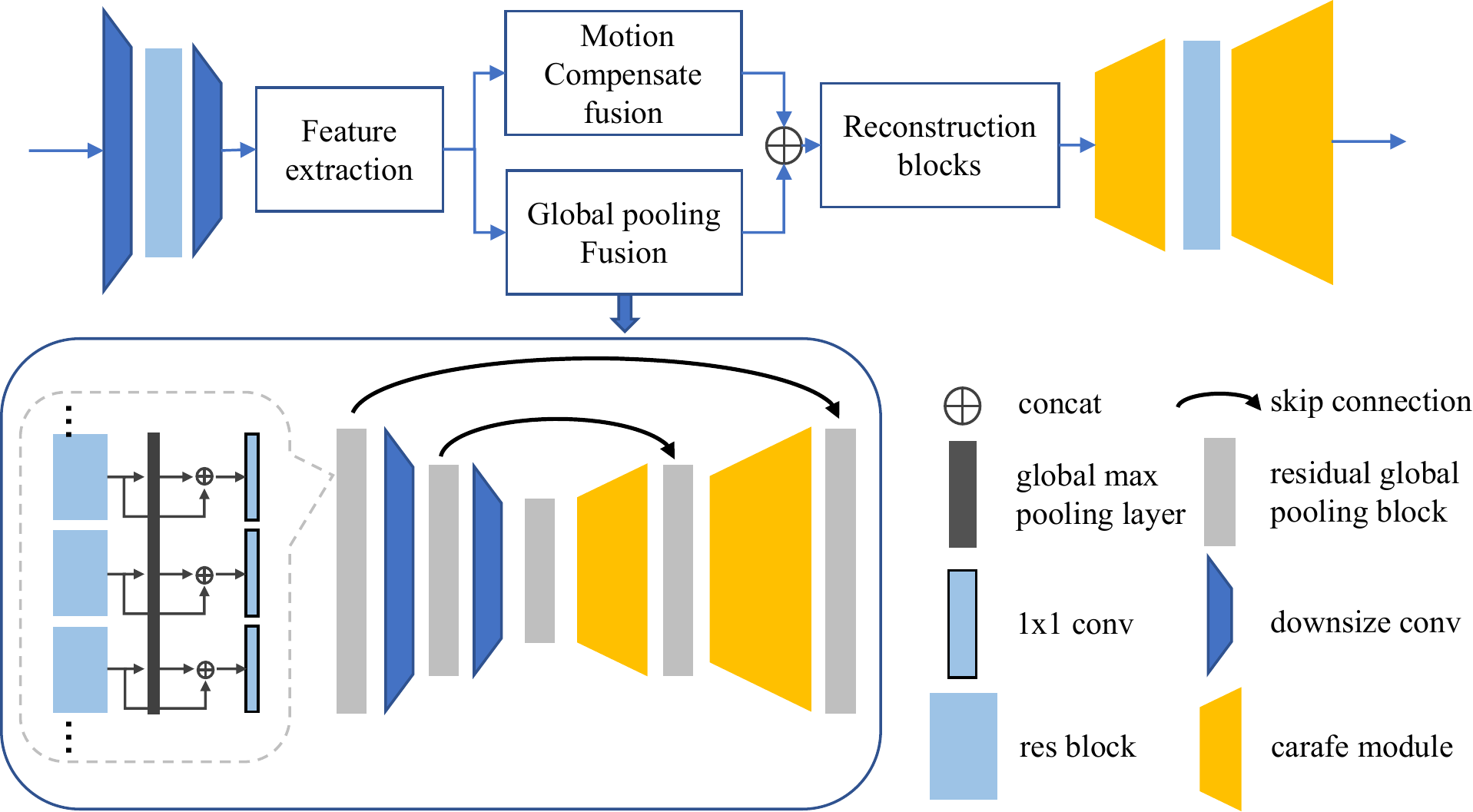}
\vspace{0mm}
\caption{Network architectures used by team JOJO-MVIG.
}
\label{fig:jojo_network}
\vspace{-1mm}
\end{figure}

\subsection{JOJO-MVIG team} 
The team proposes a unified dual-path model to jointly utilize spatial and temporal information and map low-quality compressed frames to high-quality ones. 
As shown in Fig.~\ref{fig:jojo_network}, the model consists of a feature extraction stage, two spatio-temporal fusion paths, and a reconstruction module. The overall design of the pipeline follows \cite{wang2019edvr}.

In the feature extraction part, the multi-level features are calculated. 
The fusion stage explores spatial and temporal correlation across input frames and fuses useful information. Two fusion paths are designed for motion compensation and global pooling.
The motion compensation fusion part measures and compensates the motion across frames by aligning them to the reference frame. The fusion is performed on aligned frames/features. The team adopts the alignment and fusion part from EDVR~\cite{wang2019edvr} for the motion compensation part.

Compared to the motion compensation path, the global pooling fusion path requires no alignment and adopts a U-net~\cite{Unet} like architecture in which global max-pooling layers are inserted into all residual blocks.
Global pooling has been used in \cite{PICN_2018_ECCV} to conduct permutation invariant deblurring. 
Here global pooling is used to exchange information between different frames, and since max-pooling is a selective process, different frames vote for the best information for restoration. Furthermore, the team adopts the CARAFE Module~\cite{CARAFE} to enable pixel-specific content-aware up-sampling.
More specifically, the team uses 7 frames as input, with reconstruction blocks consisting of 40 residual blocks and feature extraction module consisting of 5 residual blocks. The channel number for each residual block is set to 128.

\begin{figure}[t]
\begin{center}
\includegraphics[width=1\linewidth]{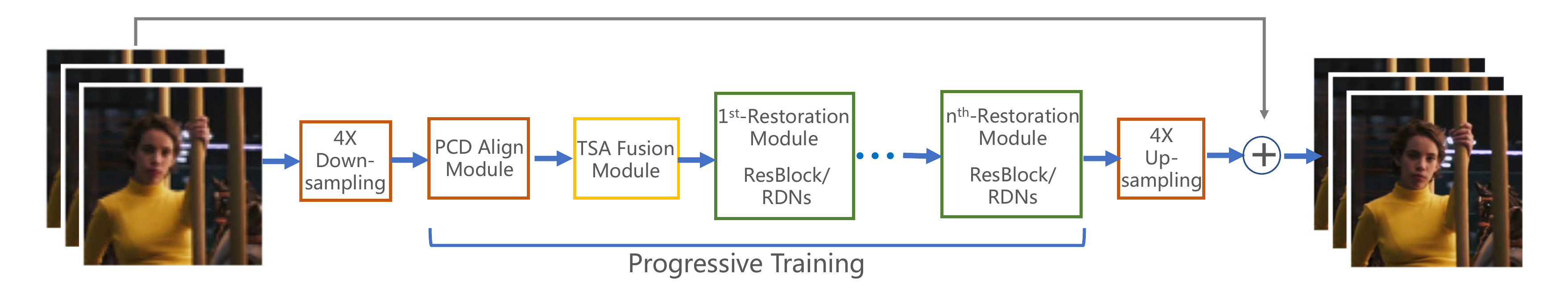}
\end{center}
   \caption{Illustration of the proposed framework by BossGao. The team exploits cutting-edge deep neural architectures for the video quality mapping task, i.e PCD align module, TSA fusion module, residual blocks and RDN blocks. For progressive training, first, the PCD align module and the 1st Restoration module are trained together. Next, the TSA fusion module is plugged in and the existing parameters are used as initialization. Then, the new framework with TSA module is trained again. More restoration modules can be stacked to get a deeper framework, which can be trained to achieve better performance.}
\label{fig:bossgao_framework}
\end{figure}

\subsection{BossGao team} 

The BossGao team exploits cutting-edge deep neural architectures for the video quality mapping task. Specifically, the team develops the following frameworks: 
\begin{itemize}
    \item Framework1: PCD+TSA+10ResBlocks+30ResBlocks
    \item Framework2: PCD+RDN1
    \item Framework3: PCD+TSA+RDN1
    \item Framework4: PCD+TSA+RDN2
\end{itemize}
where 10 ResBlock means 10 residual blocks~\cite{edsr}, and there are two convolution layers in each ResBlock.
RDN1 denotes 10 RDBs~\cite{rdn} with 8 convolution layers in each RDN.
RDN2 denotes 8 RDBs with 6 convolution layers in each RDN.
PCD and TSA are proposed in~\cite{edvr}. The framework is illustrated in Fig.~\ref{fig:bossgao_framework}.

Another contribution of the team is that, they propose to train the modules in these frameworks progressively. They train a framework by starting with fewer modules. More modules are added in progressively. When new modules are plugged in, the existing parameters are used as initialization, and the new modules and old modules are trained together. The modules in their frameworks are added in a carefully arranged order. Specifically, a framework with a PCD module and shallower restoration modules is trained first. Then, a TSA module is plugged in. Furthermore, more restoration modules can be stacked on to get a deeper frameworks. Frameworks trained by their method achieve better performance than the corresponding networks that are trained once-off.

In the final phase, the frameworks with the best performance are selected to produce the final test videos, i.e. Framework1, Framework3 and Framework4. Framework2 is only used for the last submission in the development phase.



\subsection{DPE (baseline for track 2)} 

DPE \cite{chen2018deep} is originally developed for weakly-supervised photo enhancement. For track 2, we apply it to enhance videos frame by frame. In particular, DPE treats the problem with a two-way GAN whose structure is similar to CycleGAN \cite{zhu2017unpaired}. To address the unstable training issue of GANs and obtain high-quality results, DPE proposes a few improvements along the way of constructing the two-way GAN. First, it suggests to augment the U-Net \cite{ronneberger2015u} with global features for the design of the generator. In addition, individual batch normalization layers are proposed for the same type of generators. For better GAN training, DPE proposes an adaptive weighting Wasserstein GAN scheme. 

\subsection{WESPE (baseline for track 2)} 

Similar to DPE \cite{chen2018deep}, WESPE \cite{ignatov2018wespe} is another baseline that exploits the GAN technique for weakly supervised per-frame enhancement. The WESPE model comprises a generator $G$ paired with an inverse generator $G_r$. 
In addition, two adversarial discriminators $D_c$ and $D_t$ and total variation (TV) complete the model's objective definition.
$D_c$ aims at distinguishing between high-quality image $y$ and enhanced image $\tilde{y}=G(x)$ based on image colors, and $D_t$ distinguishes between $y$ and $\tilde{y}$ based on image texture.
More specially, the objective of WESPE consists of:
i) content consistency loss to ensure $G$ preserves $x$'s content,
ii) two adversarial losses ensuring generated images $\tilde{y}$ lie in the target domain $Y$: a color loss and a texture loss,
and iii) TV loss to regularize towards smoother results.

\subsection{DACAL (baseline for track 2)} 

For track 2, we suggest the DACAL method \cite{huang2019divide} as the last baseline, which enhances videos directly. To further reduce the problem complexity, DACAL decomposes the photo enhancement process into multiple sub-problems. On the top level, a perception-based division is suggested to learn additive and multiplicative components, required to translate a low-quality image or video into its high-quality counterpart. On the intermediate level, a frequency-based division is exploited in the GAN context to learn the low- and high-frequency based distribution separately in a weakly-supervised manner. On the bottom level, a dimension-based division is suggested to factorize high-dimensional distributions into multiple one-dimensional marginal distributions for better training on the GAN model. To better deal with the temporal consistency of the enhancement, DACAL introduces an efficient recurrent design of the GAN model. 

\begin{table*}[t]
	\centering
	\newcommand{\sep}{~~}
		\begin{tabular}{@{}ll@{\sep}|@{\sep}c@{\sep}c@{\sep}c@{\sep}c@{\sep}c@{\sep}@{\sep}c@{\sep}c@{\sep}c c@{}}
			& Method  & $\uparrow$PSNR & $\uparrow$SSIM & $\downarrow$LPIPS & TrainingReq  & TrainingTime & TestReq & TestTime & Parameters & ExtraData\\
			\hline
			\multirow{7}{1mm}{\rotatebox{90}{\resizebox{14mm}{!}{Participants}}}
			& BossGao  & \textbf{32.419} & \textbf{0.905} & \underline{0.177} & 8$\times$V100 & 5-10d &  1$\times$V100 &  4s & n/a  &  No\\
			& JOJO-MVIG  & \underline{32.167} & \underline{0.901} & \underline{0.182} & 2$\times$1080Ti &  $\approx$ 4d &  1$\times$1080Ti &  2.07s & $\approx$22.75M  & No \\
			& GTQ  & \underline{32.126} & \underline{0.900} & 0.187 & 2$\times$2080Ti & $\approx$ 5d & 1$\times$2080Ti & 9.74s & 19.76M & No \\
			& ECNU  & 31.719 & 0.896 & 0.198 & 2$\times$1080Ti & 2-3d & 1$\times$1080Ti & 1.1s & n/a & No \\
			& TCL & 31.701 & 0.897 & 0.193 & 2$\times$1080Ti &  $\approx$ 3d & 1$\times$1080Ti & 25s &  $\approx$8.92M & No \\
			& GIL  & 31.579 & 0.894 & 0.195 & 1$\times$970Ti &  $\approx$ 6d & 1$\times$970Ti  &  11.37s &  3.60M &  No \\
			& $7$-th team & 30.598 & 0.878 & \textbf{0.176} & n/a & 4d & n/a & 0.5s & $\approx$7.92M & Yes \\
			\hline
		
 			\multirow{1}{1mm}
			& No processing  & 30.553 &  0.877 & \textbf{0.176} & & & & & & \\
            
			\hline
		\end{tabular}
	\caption{Quantitative results for Track 1. \textbf{Bold}: best, \underline{Underline}: second and third best. TrainingTime: days, TestTime: seconds per frame.}
	\label{tab:quantitative-results}
\end{table*}

\section{Challenge Result Analysis}

\begin{figure*}[t]
    \centering
	\includegraphics[width=0.9\textwidth]{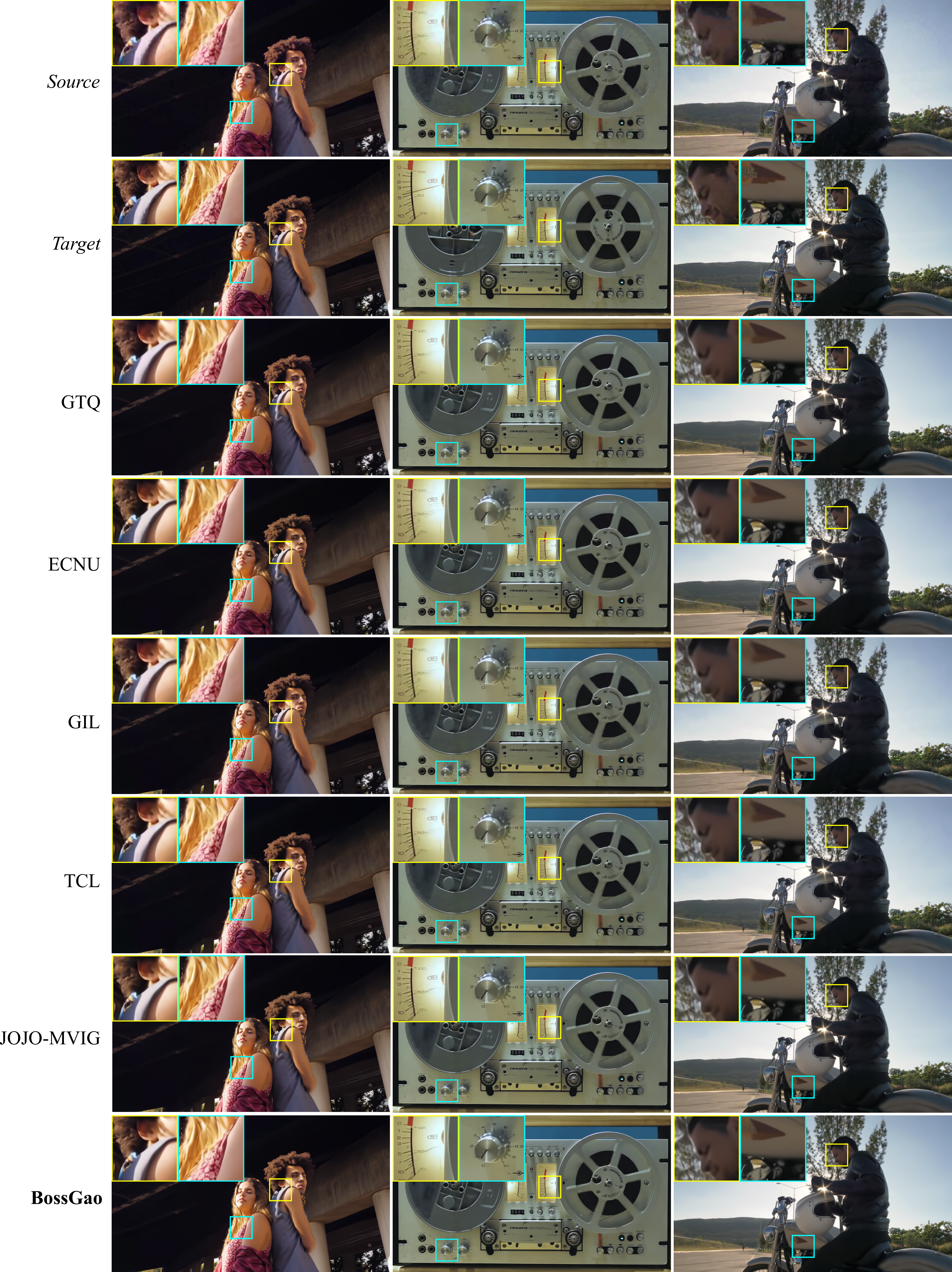}%
	\caption{Visual Comparison for Track 1.}%
	\label{fig:visual_comparison_track1}\vspace{1mm}
\end{figure*}

\begin{figure*}[t]
    \centering
	\includegraphics[width=\textwidth]{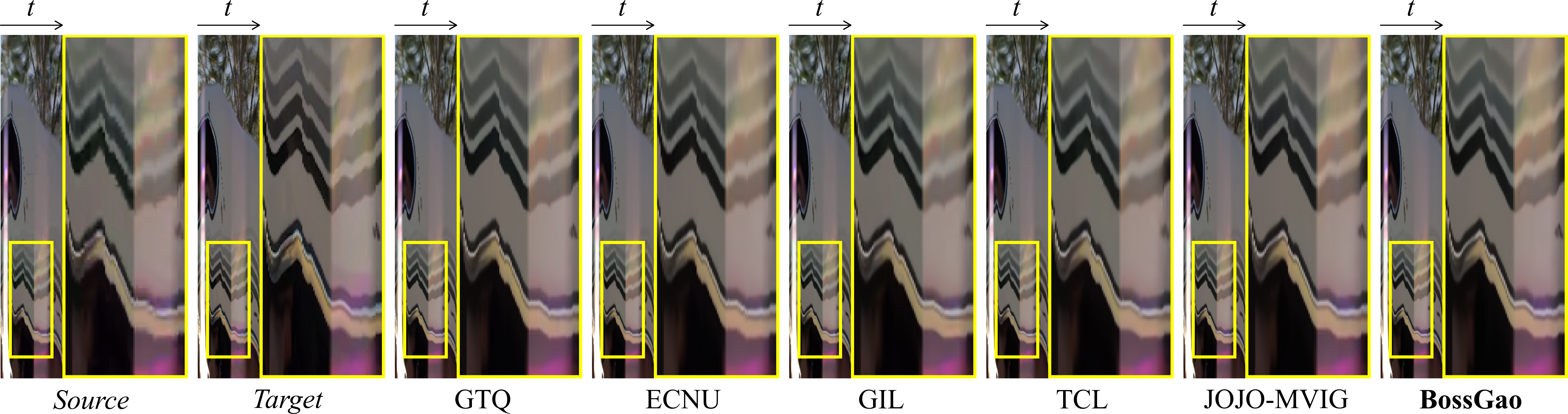}%
	\caption{Temporal Profiles for Track 1.}%
	\label{fig:temporal_profiles_track1}\vspace{1mm}
\end{figure*}

\begin{figure*}[t]
    \centering
	\includegraphics[width=0.96\textwidth]{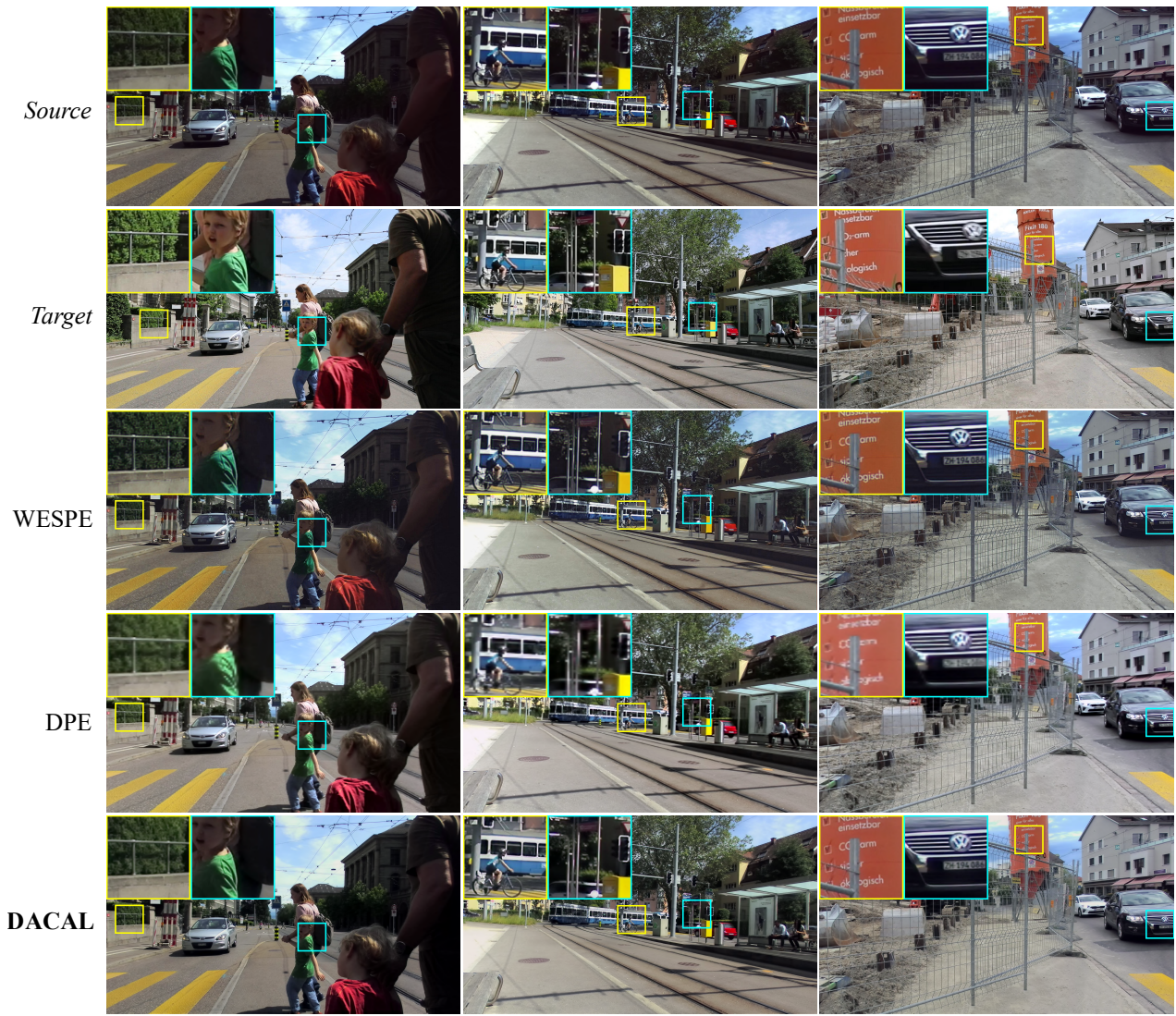}%
	\caption{Visual Comparison for Track 2.}%
	\label{fig:visual_comparison_track2}\vspace{1mm}
\end{figure*}

\begin{figure*}[t]
    \centering
	\includegraphics[width=0.625\textwidth]{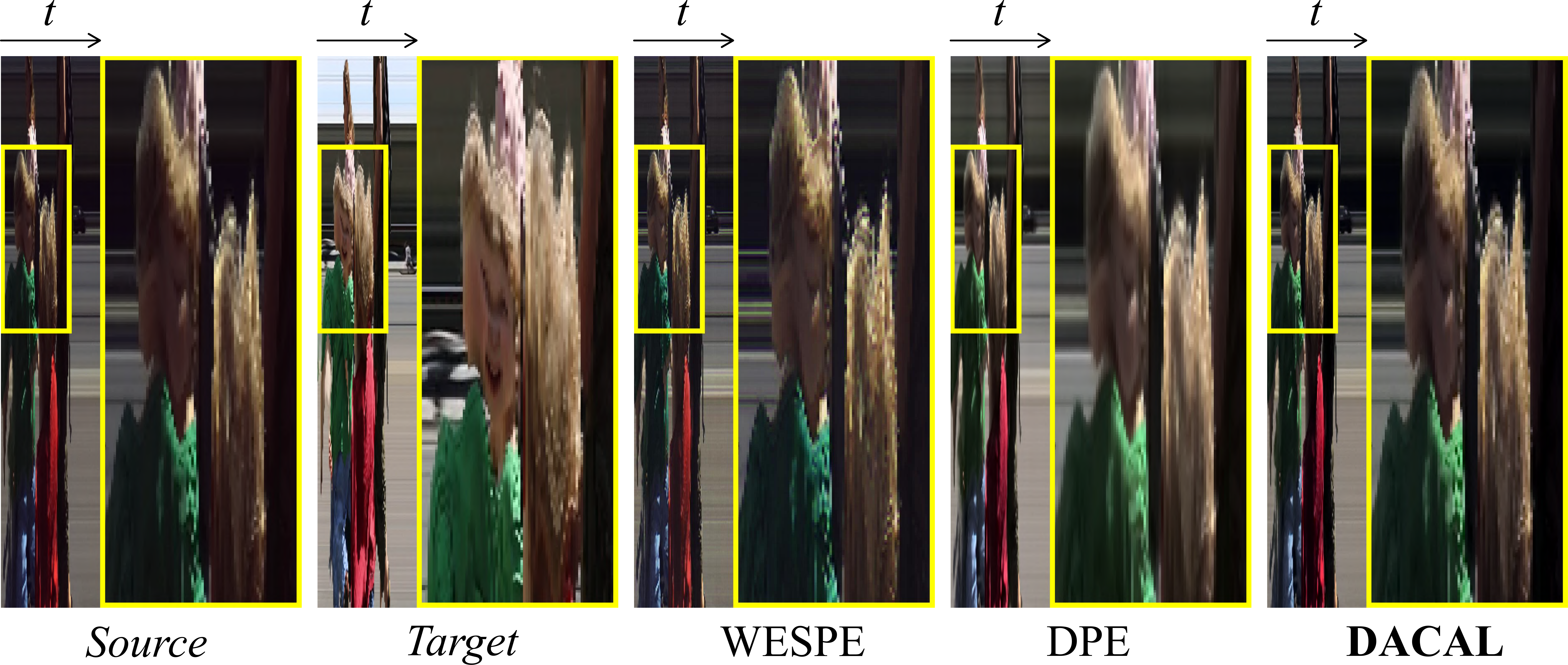}%
	\caption{Temporal Profiles for Track 2.}%
	\label{fig:temporal_profiles_track2}\vspace{1mm}
\end{figure*}

\subsection{Track 1: Supervised VQM}
This challenge track aims at restoring the discarded information, which has been lost due to compression, with the highest fidelity to the ground truth. Because of full supervision, the ranking among the participating teams can be computed objectively. 

\textbf{Metrics} The most popular full reference metrics to evaluate the quality of images and videos are PSNR and SSIM. PSNR can be computed directly from the mean-squared-error (MSE). Therefore, L2-norm based objectives are commonly used to obtain high PSNR scores. SSIM is calculated from windows based statistics in images. In this challenge, both metrics are calculated per frame and averaged over all sequences. 
Table \ref{tab:quantitative-results} reports the quantitative results of participating methods as well as the baseline, i.e. the input without any processing.
With a PSNR value of 32.42dB and SSIM score of 0.91, team BossGao achieves the highest scores overall and is the winner of challenge track 1. Team JOJO-MVIG and GTQ follow closely with a PSNR difference of 0.25dB and 0.29dB to the winner respectively. The remaining teams also achieve respectable PSNR scores slightly below 32dB. The ranking in terms of SSIM is almost the same. In addition, as can be seen by the reported training times, capacity and test times, models with more parameters and teams with more processing power generally perform better. However, team ECNU manages to surpass more expensive methods with the fastest runtime. Team GIL targets for a compact network with the least parameters, which can be trained on a single lower-end GPU but still produces promising enhancement results.

\textbf{Visual Comparison}
Selected samples from the test data are provided in Fig.~\ref{fig:visual_comparison_track1} to compare the visual quality of the enhanced video frames among all teams. The visual comparison shows that team BossGao also performs the best for the quality enhancement on such sampled frames. It should be noted that due to the inherent loss of information after compression, fidelity based methods are not able reconstruct all high frequency details and tend to over-smooth the content.
In order to assess continuity between frames, temporal profiles for all teams are provided in Fig.~\ref{fig:temporal_profiles_track1}. A single vertical line of pixels is recorded over all frames in the sequence and stacked horizontally.

Additionally, we computed LPIPS~\cite{zhang2018perceptual} scores to compare perceptual quality among the teams. Optimizing for perceptual quality was not required by the participants in this challenge track, but the metric still provides interesting insights into quantitative quality assessment and its limitations. The scores among all teams is roughly consistent with PSNR and SSIM, which implies that the top teams also produce visually more pleasing results compared to their competitors. Interestingly, the input without processing along with team 7, which basically doesn't alter the input, achieves the best score. We assume that the distortions, due to smoothing of L2-norm based methods, cause worse scores for the top teams, despite much higher reconstruction quality. In contrast, compression algorithms are designed to optimize for perceptual quality, which could lead to the strong LPIPS score for the input.

\subsection{Track 2: Weakly-Supervised VQM}
In this challenge track, the goal of the task is to enhance the video characteristics from a low quality device (ZED camera) to the characteristics of a high-end device (Canon 5D Mark IV) with limited supervision. Weak supervision is provided by weakly-paired videos, which share approximately the same content and are roughly aligned in the spatial and temporal domain.

\textbf{Metrics} Since there is no pixel-aligned ground truth available, full reference metrics are no option for quality assessment. Usually, results for these types of problems are scored by a MOS study, conducted by humans visually comparing different methods. While there exist metrics to measure distances between probability distributions for high level content, e.g. Fr\'{e}chet Inception Distance (FID)~\cite{fid}, that are widely applied to generative models, finding reliable metrics for low-level characteristics remains an open problem. 
Popular perceptual metrics such as Learned Perceptual Image Patch Similarity ~\cite{zhang2018perceptual} metric and Perceptual Index~\cite{blau20182018} are used in the field too. However, we found the scores for these metrics are not suitable for the problem setting in this challenge and do not always correlate with human perception. Perceptual Index is not a relative score, it only measures general quality. However, we are interested in measuring the mapping quality from one domain to another. LPIPS requires aligned frames which is a problem since the frames are only roughly aligned. Nevertheless, we provide LPIPS scores for a selection of methods along with visual results, see Table~\ref{tab:lpips_track2}, Fig.~\ref{fig:visual_comparison_track2} and Fig.~\ref{fig:temporal_profiles_track2}. Surprisingly, the source without processing achieves comparable scores to the enhancement methods. While source and target frames are captured by a real camera, the methods alter the videos artificially. Since LPIPS relies on a feature extractor, which is trained on real images, this could lead to unsatisfactory scores for the methods, due to low level distortions.

\begin{table}[t]
	\centering
	\newcommand{\sep}{~~}
		\begin{tabular}{@{}l|@{\sep}@{\sep}c@{\sep}c@{\sep}c@{\sep}c@{\sep}c@{\sep}@{\sep}c@{\sep}c@{\sep}c c@{}}
			& Source & WESPE & DPE & DACAL\\
			\hline
            LPIPS$\downarrow$ & 0.590 & \underline{0.584} & 0.631 & \textbf{0.565} \\
			\hline
		\end{tabular}
	\caption{LPIPS scores for Track 2. \textbf{Bold}: best, \underline{Underline}: second best.}
	\label{tab:lpips_track2}
\end{table}

\textbf{Visual Comparison}
Since there are no submissions for this track, visual results and temporal profiles for a selection of recent image and video quality mapping methods is provided as reference in Fig.~\ref{fig:visual_comparison_track2} and Fig.~\ref{fig:temporal_profiles_track2} . WESPE~\cite{ignatov2018wespe} and DPE~\cite{chen2018deep} are single image methods which are applied per frame, DACAL~\cite{huang2019divide} is a true video enhancer. All the competing methods are trained on the Vid3oC  dataset~\cite{kim2019vid3oc}. The visual results show that DACAL preserves more details and enhances contrast better, while WESPE introduces biased colorization and DPE produces blurry textures. 

	\section{Conclusions}
This paper presents the setup and results of the NTIRE 2020 challenge on video quality mapping. This challenge addresses two real world settings: track 1 concerns video quality mapping from more compressed videos to less compressed ones with available paired training data; track 2 focuses on video quality mapping from a lower-end device to a higher-end device, given a collected weakly-paired training set. 7 teams competed in Track 1 in total. The participating methods demonstrated interesting and innovative solutions to the supervised quality mapping on compressed videos. In contrast, we evaluated three existing methods for track 2, showing their performance is promising but much effort is still needed for better video enhancement. 
The evaluation with LPIPS on both challenge tracks reveals the limits of current quantitative perceptual quality metrics and shows the need for more research in that area, especially for track 2 where no pixel-aligned reference is available.
Our goal is that this challenge stimulates future research in the area of video quality mapping in either  supervised or weakly-supervised scenarios, by serving as a standard benchmark and by the evaluation of new baseline methods.
	\section*{Acknowledgements}
We thank the NTIRE 2020 sponsors: Huawei, Oppo, Voyage81, MediaTek, DisneyResearch$\mid$Studios, and Computer Vision Lab (CVL) ETH Zurich.

	\section*{Appendix A: Teams and affiliations}
\label{sec:affiliation}

\subsection*{NTIRE 2020 VQM organizers}
\textbf{Members:} Dario Fuoli (dario.fuoli@vision.ee.ethz.ch), Zhiwu Huang (zhiwu.huang@vision.ee.ethz.ch), Martin Danelljan (martin.danelljan@vision.ee.ethz.ch), Radu Timofte (radu.timofte@vision.ee.ethz.ch)

\textbf{Affiliations:} Computer Vision Lab, ETH Zurich, Switzerland

\subsection*{GTQ team}

\textbf{Title:} Modified Deformable Convolution Network for Video Quality Mapping

\textbf{Members:} Hua Wang, Longcun Jin (lcjin@scut.edu.cn), Dewei Su

\textbf{Affiliations:} School of Software Engineering, South China University of Technology, Guangdong,
China

\subsection*{ECNU team}
\textbf{Title:} Compression2Compression: Learning compression artifacts reduction without clean data

\textbf{Members:} Jing Liu (splinter02@163.com)

\textbf{Affiliations:} Multimedia and Computer Vision
Lab, East China Normal University, Shanghai, China

\subsection*{GIL team}
\textbf{Title:} Dense Recurrent Residual U-Net for Video Quality Mapping

\textbf{Members:} Jaehoon Lee (dlwogns0729@gmail.com)

\textbf{Affiliations:} Department of Electronics and Computer Engineering, Hanyang University, Seoul, Korea

\subsection*{TCL team}
\textbf{Title:} Deformable convolution based multi-frame network with single-frame U-Net

\textbf{Members:} Michal Kudelski (michal.kudelski@tcl.com),  Lukasz Bala, Dmitry Hrybov, Marcin Mozejko

\textbf{Affiliations:} TCL Research Europe, Warsaw, Poland

\subsection*{JOJO-MVIG team}
\textbf{Title:} Video Quality Mapping with Dual Path Fusion Network

\textbf{Members:} : Muchen Li (muchenli1997@gmail.com), Siyao Li, Bo Pang, Cewu Lu

\textbf{Affiliations:} Machine Vision and Intelligence Group, Department of Computer Science, Shanghai Jiao Tong University, Shanghai, China

\subsection*{BossGao team}
\textbf{Title:} Exploiting Deep Neural Architectures by Progressive Training for Video Quality Mapping

\textbf{Members:} Chao Li (uqcli1@gmail.com), Dongliang He, Fu Li, Shilei Wen

\textbf{Affiliations:} Department of Computer Vision Technology (VIS), Baidu Inc., Beijing, China

	{\small
		\bibliographystyle{ieee_fullname}
		\bibliography{egbib}
	}
	
\end{document}